# Continuous Dynamic Photostimulation - inducing in-vivo-like fluctuating conductances with Channelrhodopsins

**Running title:** Continuous dynamic photostimulation by ChIEF-activation

Andreas Neef [1,3,4]*, Ahmed El Hady[1,2,3,4]*, Jatin Nagpal[1,3,4]*, Kai Bröking[1], Ghazaleh Afshar[1,3,4], Oliver M Schlüter [5], Theo Geisel[1,3,4], Ernst Bamberg[6,3], Ragnar Fleischmann[1], Walter Stühmer[2,3,4], Fred Wolf [1,3,4]

[1]Max-Planck-Institute for Dynamics and Self-Organization, Göttingen, Germany
[2]Max-Planck-Institute for Experimental Medicine, Göttingen, Germany
[3]Bernstein Focus Neurotechnology, Göttingen, Germany
[4]Bernstein Center for Computational Neuroscience, Göttingen, Germany
[5]European Neuroscience Institute, Göttingen, Germany
[6]Max-Planck-Institute for Biophysics, Frankfurt/Main, Germany
* these authors contributed equally

Corresponding authors:   Andreas Neef,   andreas@nld.ds.mpg.de,   +49 551 5176 414
Fred Wolf,   fred@nld.ds.mpg.de,   +49 551 5176 419

Word count (intro, Results, Discussion): 3,645 (soft max: 3,500); Abstract: 145

## Abstract

Central neurons operate in a regime of constantly fluctuating conductances, induced by thousands of presynaptic cells. Channelrhodopsins have been almost exclusively used to imprint a fixed spike pattern by sequences of brief depolarizations. Here we introduce continuous dynamic photostimulation (CoDyPs), a novel approach to mimic *in-vivo* like input fluctuations noninvasively in cells transfected with the weakly inactivating channelrhodopsin variant ChIEF. Even during long-term experiments, cultured neurons subjected to CoDyPs generate seemingly random, but reproducible spike patterns. In voltage clamped cells CoDyPs induced highly reproducible current waveforms that could be precisely predicted from the light-conductance transfer function of ChIEF. CoDyPs can replace the conventional, flash-evoked imprinting of spike patterns in *in*-vivo and *in*-vitro studies, preserving natural activity. When combined with non-invasive spike-detection, CoDyPs allows the acquisition of order of magnitudes larger data sets than previously possible, for studies of dynamical response properties of many *individual* neurons.

# Introduction

Channelrhodopsin 2 (ChR2) and other optogenetic tools promise to revolutionize the way neuroscience is done by offering a non-invasive control of neuronal activity with high spatial and temporal resolution[1,2]. Optogenetic tools have been proposed and implemented for advancing the analysis of neuronal systems at all levels from single cells through circuits' structure and function[3-5] up to the level of behaviour[6]. Recently ChR2 was also utilized for non-invasive stimulation in cellular and synaptic physiology[7,8]. The development of novel channelrhodopsin-variants with advantageous properties like sustained activation after brief illumination[9], reduced inactivation[1] or neuronal excitation at very low light levels[10] are extending the applicability of channelrhodopsins to problems for which previously the demands on light-power density or stimulation frequency or the extent of induced depolarization could not be met. However, so far the use of ChR2 has been restricted to pulsed[11], constant or slowly ramped[12] illumination in order to produce precisely timed spikes following the light stimulation or in order to induce oscillations *in-vivo*. To date, there was no study that attempted to attain continuous control of conductance using ChR2 or any of its variants in order to emulate *in-vivo* like subthreshold fluctuations.

A typical neuron, embedded in a cortical network *in-vivo*, receives about 10.000 synaptic inputs[13]. Assuming that each of these synaptic inputs is active with a rate on the order of 1 to 10 Hz, incoming signals arrive at a rate of 10 kilohertz (Fig. 1a). As a result, the membrane voltage exhibits strong, temporally irregular fluctuations. To understand the computational capabilities of e.g. cortical circuits, it is essential to characterize single neuron computation under such realistic operating conditions. To control the activity of entire neuronal circuits while preserving their natural firing characteristics, it would be advantageous to introduce artificial input components mimicking intrinsically generated synaptic input but under precise experimental control. While optogenetic tools have been successfully used to manipulate activity patterns in intact neuronal circuits, the strong light flashes typically used completely override intrinsic activity and rigidly imprint artificial spike sequences. Those studies of single neuron computation that preserve natural operation conditions[14-19] so far rely on invasive stimulation methods (Fig. 1b) that are severely constrained by limited recording duration and low yield. Combining noninvasive optogenetic stimulation with multichannel multi-neuronal recordings promises to overcome these limitations (Fig. 1c). The characterization of single neuron computation requires a precise knowledge of the input in order to compute for instance the spike triggered average or covariance of the stimulus or to describe correlation gain and firing rate adaptation in dependence of the stimulus properties. An optical, noninvasive stimulation approach is only instrumental for such studies if:

1. The induced conductances are highly reproducible with correlation times suitable to mimic fluctuating synaptic conductances

2. Waveforms can be precisely designed and predicted

3. Conductance waveforms can be stably induced in long-term experiments

Satisfying these requirements would also provide the basis to control the activity of operating circuits *in-vivo*, preserving the natural firing characteristics. The possibility for long-term illuminations at biologically tolerated light intensities came with the development of weakly inactivating Channelrhodopsin variants.

In this study, we introduce continuous dynamic photostimulation CoDyPs mimicking *in-vivo*-like fluctuations using light gated ion channels (Fig.1). We assess the feasibility of CoDyPs, by mapping the fluctuations in the light-power density to fluctuations in the light induced currents and by a characterization of the reproducibility of the induced currents. We test whether the photocurrents can be predicted from the fluctuating light pulses using linear system theory and extend the use of CoDyPs to drive action potentials in cultured neurons by *in-vivo* like fluctuations for several hours. We present here evidence that ChIEF and, to a lesser degree, ChR2 are suited to induce a well defined, temporally fluctuating conductance. Both channels respond to the blue light stimulus similar to a low-pass filter with a cut-off frequency of 20 Hz. The current responses were reproducible and could be predicted by convolution of the channel's impulse response function with the fluctuating light input. In summary, ChIEF meets the three criteria posed above and seems suitable for long-term studies of dynamic response properties of cortical neurons by CoDyPs. To our knowledge, this is the first study to investigate the use of ChR2 and its variants in mimicking *in-vivo* like subthreshold fluctuations.

## Results

Before light sensitive channels can be used as fluctuating conductances in neurons, the time-course of the induced conductances is best studied in cells with little background conductances to characterize the light-induced conductance in isolation. To this end we used Human embryonic kidney cells 293 (HEK cells). To study ChR2, we created a stable ChR2-YFP cell-line, ChIEF was studied in HEK cells transiently transfected with ChIEF-tdTomato. HEK cells were chosen for patch clamp by their appearance in bright field and their fluorescence intensity. Under whole cell voltage clamp, with the membrane potential fixed at -60 mV, light pulses of different power density were applied. The elicited currents exhibited the typical features of ChR2 and ChIEF currents (Fig. 2a) similar to what has been reported before[1]: a rapid activation at light onset with activation time constant $\tau_{act}$ = 4.8±0.1 ms for ChR2, $\tau_{act}$ = 4.8± 0.2 ms for ChIEF (Fig. 2c and Methods) and a rapid deactivation after cessation of illumination ($\tau_{1deact}$ = 8.5±0.9 ms for ChR2, $\tau_{2deact}$ = 30±7 ms for ChR2 , $\tau_{1deact}$ = 6.9±0.4 ms for ChIEF, $\tau_{2deact}$ = 66±15 ms for ChIEF ) as well as a slower inactivation after an initial peak ($\tau_{inact}$ =63±2 ms for ChR2, $\tau_{inact}$ = 185±16ms for ChIEF). While the deactivation kinetics and the light-dependent activation time constant were similar for ChR2 and ChIEF, the inactivation kinetics and especially the degree of inactivation were substantially different. At the maximum light-power density used in this study, 0.27 mW/mm$^2$, ChR2 currents showed an inactivation of 58.8 % ± 0.8 %, resulting in an average current steady state current of only 57±11 pA. The new mutant ChIEF, on the other hand, showed only 13.4 % ± 0.9 % inactivation from the peak to steady state level with a steady state level of 280 ± 68 pA. Reported values are mean ± standard error for 13 cells (ChR2) and 21 cells (ChIEF) respectively.

To mimic naturally occurring input fluctuations, light sensitive channels must provide conductance changes with appropriate magnitude and frequency bandwidth. Thus, we first analyzed the bandwidth of currents mediated by ChIEF and ChR2. A simple, practical test is the application of chirps: over a 3 s period a pseudo-periodic light stimulus, frequency-modulated continuously from 5 to 100 Hz, was applied. While the modulation depth of the stimulus amplitude is constant, the amplitude of the current response decreases as the frequency exceeds the bandwidth of the channel. To avoid a contamination of this amplitude decrease, by the rather slow time and light-dependent inactivation of ChIEF the chirps were preceded by a 3 s constant light stimulus in the case of ChIEF. The photoactivated currents induced by the chirp light stimuli were very similar for ChR2 and ChIEF (Fig. 2d). To compare the effect of the signal transduction with a single-pole low-pass filter, the chirp stimuli were digitally filtered (see methods) and the power spectral density of the results was compared with the power spectral density of the currents. The best match was obtained with a cut-off frequency of 20 Hz (Fig. 2e), corresponding to time constants around 8 ms. This performance is sufficient to synthesize the fluctuating inputs originating from AMPA, NMDA and GABA mediated synaptic currents; the white noise limit cannot be implemented by these means. ChIEF and ChR2 appear to enable the generation of fluctuating currents with high frequencies to the same degree. The main difference between the two channelrhodopsin variants is the larger amplitude of ChIEF-mediated currents (Fig. 2b), in part due to the strongly reduced inactivation. The larger current amplitude constitutes an advantage *per se*, enabling adequate depolarization and larger fluctuations with less average light-power density. Thereby, ChIEF currents promise better reproducibility of fluctuating currents because high frequency stimulus components are not drowning in the noise floor, as it is the case for ChR2 currents (Fig. 2e). Consequently the analysis of fluctuating currents was performed in HEK cells expressing ChIEF.

*Highly reproducible fluctuating light-activated currents*

Because the drumfire of exponentially decaying postsynaptic currents can be well approximated with Ornstein Uhlenbeck (OU) processes, OU currents have been extensively used to emulate the temporally fluctuating input currents of cortical neurons *in-vivo*. To directly examine whether fluctuating inputs can be reliably imposed by photostimulation, we subjected ChIEF expressing HEK-cells to fluctuating light stimuli synthesized from an OU process. In total, 12 different stimulus ensembles were used, comprising three different correlation times $\tau_{corr}$ = 1, 5 and 50 ms and four different light-power density statistics (conditions c1 to c4, see methods). Assessing the stability and reproducibility of the induced currents, we found the trial to trial variations very small and the photoactivated currents induced by identical, successive stimuli very similar. Correlation coefficients typically ranged from 0.9 to 0.99 for correlation times of 5 ms and 50 ms (Fig. 3c) and the deviations of individual currents from the average current were generally below 5 pA (orange and black traces in Fig. 3a). We then examined how the amplitude and frequency content of the current signal were influenced by the stimulus parameters. For fluctuations with a correlation time of 50 ms - larger than the activation and deactivation time constants of the channelrhodopsin - the induced current largely mirrors the stimulus (Fig. 3a, left panel), essentially following the steady-state relation between current and light density (Fig. 2b). The probability density function of the current amplitude was very similar to that of the stimulus (Fig. 3b). For faster fluctuating stimuli with correlation times of 5 ms and 1 ms channelrhodopsin's gating kinetics limits the frequency spectrum of the current response such that the amplitude spectrum is of lower bandwidth than the stimulus (not shown) and the current amplitude distribution becomes narrower (figure 3b middle and right panel).

*Channelrhodopsin acts as a low pass filter*

The characterization of dynamical response properties of neurons under fluctuating current input requires detailed knowledge of the individual applied current waveform. In invasive approaches this waveform is directly available. In a non-invasive photo-stimulation approach the current's statistics and the time course has to be obtained computationally from the light stimulus alone. To further study the relation between stimulus and current change, we calculated the average autocorrelation function and the average impulse response function for each of the three correlation times and the four combinations of mean and standard deviation used (conditions c1 to c4, see Methods).

While the autocorrelation functions of the light stimuli decayed exponentially by construction, the autocorrelation functions of the currents fell off slower. They were well described by the auto-correlation function of an Ornstein-Uhlenbeck process with the correlation time $\tau_{corr}$ (i.e. 1, 5 or 50 ms) passed through a first order low-pass filter (equation 3) with time constant $\tau_{cut-off}$ (Fig. 4a). When the correlation time $\tau_{corr}$=50 ms is much larger than $\tau_{cut-off}$, the shape of the autocorrelation function is hardly influenced by the filtering and consequently the estimates of $\tau_{cut-off}$ do vary between 6.7 ms and 10 ms. For the smaller correlation times however, $\tau_{cut-off}$ dominates the shape of the autocorrelation function and can be well estimated. It was found to depend only weakly on the stimulus parameters, increasing slightly from 8 ms to 9 ms with decreasing mean light-power density. This relation most likely reflects the dependence of the activation kinetics on the light-power density (Fig. 2c).

Since these results are consistent with a simple linear filter model of the relation between the light stimulus and the induced currents, we wanted to examine, whether the current waveforms could be predicted by convolution of the impulse response function with the light waveform. To this end impulse response functions were estimated by inverse Fourier-transformation of the ratio between the Fourier-transforms of current response and respective light stimulus (see Methods). As expected, the standard error of the estimated impulse response functions increased with the square root of $\tau_{corr}$ (Fig. 4b). In addition, the noise increased slightly with decreasing light amplitude and standard deviation of the light stimulus (not shown).

At a membrane potential of -60 mV, at which the fluctuating light stimuli were applied, the photoactivated current is inward and so the main component of the impulse response function is negative too. Initially, however, it starts with a very brief transient of positive amplitude (Fig. 4b, inset in right panel). At the sampling frequency of 10 kHz this transient is represented by a single sample point that appeared consistently in all experiments. This transient decrease in light activated current immediately after an increase in light-power density is reminiscent of the transient response to a 10 ns flash of green light[20] (544 nm). After this brief initial transient, the impulse response function resembles that of a low pass filter: a very rapid onset followed by a single exponential decay (Fig. 4b). Again, as was the case for the autocorrelation, the time constant of this decay was only weakly stimulus dependent and decreased with increasing light-power density, from 9 ms at 0.108 mW/mm² to 7.5 ms 0.162 mW/mm². The decay phase of the impulse response function represents the effective rate with which the protein reacts to the light fluctuations around the average light intensity, distinct from activation or inactivation kinetics. This effective rate is a mixture of activation and deactivation rates and therefore increases with light. Different from a simple low-pass filter, the impulse response function of ChIEF has a delay of about 200-300 µs. It presumably resembles the transition from state P1 to P2 [20], the conformational switch after absorption of a photon.

*Computational reconstruction of conductance fluctuations*

The response of a time-invariant, linear system is fully determined by the stimulus and the impulse response function. We thus tested the predictive power of the impulse response function by convolution with the light stimulus and found that the current waveform predicted in this fashion and the average recorded current were highly congruent (Fig. 5). The mean correlation coefficients between predicted and recorded currents were, across all conditions c1 to c4, 0.976±0.002 for $\tau_{corr}$=1 ms, 0.98±0.0007 for $\tau_{corr}$=5 ms and 0.973±0.0007 for $\tau_{corr}$=50 ms (mean ± standard error, see also Fig. 5 c). Predictability was lowest for cells with a smaller trial-to-trial correlation coefficient of individual current responses, indicating that light-induced currents are indeed predicted very well and that prediction performance was limited only by noise introduced by other conductances such as leak.

*Estimation of proton fluxes*

Different from previous applications with episodic stimulation by light activated conductances, the continuous stimulation for extended periods of time will result in a large total charge transfer. To assess the ionic composition of this massive ion flux and especially the potentially harmful proton flux we used the Goldman-Hodgkin-Katz flux-equation. ChR2 is mainly permeable for sodium/potassium and protons, the ratio of the two permeabilities is about $P_{H+}/P_{Na/K}=10^6$ [2]. The chimera ChEF, which differs from ChIEF by a single point mutation and likely has the same pore permeabilities as ChIEF[1], has permeability ratios $P_{H+}/P_{Na+}=0.877\times10^6$ and $P_{K+}/P_{Na+}=0.673$ [1]. These values allow an approximation of the fractional currents flowing during CoDyPs assuming standard physiological concentrations of the permeating ions (see Methods): the flux ratios are $\Phi_K/\Phi_{Na}=$ -0.05 and $\Phi_H/\Phi_{Na}=0.53$. If potassium, sodium and proton fluxes are assumed independent of each other, 36% of the current amplitude are carried by protons; this large fraction is not surprising as ChR2 is a leaky proton pump[21]. Although the independence of fluxes might not apply, this value can serve as a preliminary estimate. For a spherical cell with diameter of 15 µm it predicts a proton influx equivalent to 0.4 mmol/l per second for a photocurrent current of 200 pA.

*Long-term CoDyPs of cultured neurons*

So far our results demonstrate that conceivably CoDyPs satisfies the requirements for a noninvasive stimulation method with respect to reproducibility, bandwidth and predictability. The estimation of the fractional proton flux, however, raises the question whether cells will be able to handle the involved proton influx. We therefore set out to directly test, whether CoDyPs is suitable for noninvasive long-lasting experiments without compromising neuron survival and most importantly neuronal response properties. To this end neurons were cultured on multielectrode arrays to detect the action potentials and transfected with Chop2. Pharmacological block of synaptic transmission abolished all spontaneous action potentials. The cells were repeatedly exposed to 60 minutes continuous stimulation with fluctuating light, interrupted by 60 minutes of darkness. In experiments lasting up to 9 hours neuronal action potential patterns were remarkably stable and systematically related to the predicted conductance input (Fig. 6b). The firing rate displayed a systematic relation to stimulus parameter ((Fig. 6a), most notably a transient. reminiscent of firing rate adaptation at the onset of each 60 minute stimulation period but also a small increase when the correlation time was increased from 5 ms to 10 ms. As action potential patterns are very sensitive to changes in the membrane potential, we conclude that the neurons' conditions were stable, indicating that CoDyPs is indeed a very valuable tool in studying dynamic properties of neurons, allowing noninvasive stimulation for many hours.

## Discussion

Controlled naturalistic stimulation of neurons and sensory systems is a powerful experimental strategy that has revealed fundamental aspects of neuronal processing including high rates of encoded sensory information[22-24] and the surprisingly broad bandwidth of cortical population dynamics[19,25-27]. Naturalistic stimulation aims to characterize neuronal dynamics under *in-vivo* like operating conditions. Theoretical neuroscience has developed and validated computational concepts and tools of steadily increasing sophistication to model and analyze neuronal operations in the fluctuation driven firing regime[15,28-37]. In the present study we have developed an optogenetic approach that meets the key requirements of an experimental approach for corresponding experimental studies: the inputs are reliable and reproducible, offer the necessary bandwidth and the stimulus waveform can be conveniently predetermined. This non-invasive, yet controlled stimulation method has the potential to revolutionize data collection in this field of neuroscience, enabling large-scale screening or targeted studies of cellular mechanisms. CoDyPs is distinct from most previous applications of excitatory optogenetic tools. Both, *in-vivo* and *in-vitro* many studies succeeded in controlling impulse activity by imprinting action potential sequences stimulating with sequences of light flashes[11] or raising firing rates by steady depolarization[38]. Some of the latest advances in the engineering of channelrhodopsins have specifically enhanced the usability of light gated ion channels for this type of applications (Cheta[39], CatCh[10], bistable ChRs[9]). In contrast CoDyPs drives cells by ongoing quantitatively controlled conductance fluctuations. In this approach the neuron "decides" whether and when to generate action potentials in a way that reflects a realistic interplay of intrinsic dynamics and complex input patterns. It is worth noticing that CoDyPs is facilitated by the slow and weak inactivation of ChIEF and its low single channel conductance, molecular features that are not specifically beneficial for precisely imprinting predetermined spiking patterns. ChIEF's strongly reduced inactivation supports the generation of ongoing conductance fluctuations around a maintained mean level. The generally small single channel conductance of Channelrhodopsins is the basis of the small trial to trial fluctuations that make CoDyPs currents highly reproducible. The molecular property that appears as the most severe limitation of currently available optogenetic tools for CoDyPs is the characteristic response time on the order of 7 to 8. It would be desirable to develop Channelrhodopsin variants with faster off-kinetics to extend the use of CoDyPs towards the white noise limit.

Perhaps the most surprising result of our study is the precision and ease with which CoDyPs induced conductance fluctuations can be predicted and designed. We found that simple linear-response theory is sufficient to computationally reconstruct dynamic conductance fluctuations with virtually perfect accuracy. In addition, filter parameters were only weakly dependent on stimulus conditions such that a small and easily parameterized library of impulse response functions appears sufficient to predict and design conductance fluctuations for a wide range of stimulation conditions. Using this approach, even in the cell with unknown ChIEF expression level, precise calibration of photon flux in the sample plane is sufficient to accurately predict the fluctuating conductance waveform. The absolute conductance scale can be adjusted using interspike interval statistics, readily obtained by extracellular recordings. In addition whole-cell recordings at the end of CoDyPs sessions would be sufficient to determine the absolute magnitude of the light induced conductance fluctuations. Together with the long term stability of

CoDyPs driven spiking patterns our findings indicate that virtually all experimental paradigms previously realized by whole cell stimulation and recording can be performed using CoDyPs. These include measurements of firing rate as a function of input statistics [17] and measurements of the dynamic gain[25-27]. With patterned illumination, each neuron can be driven by a specific stimulus, extending the use of CoDyPs to the simulation of partially correlated inputs. In this way correlations in the spike trains of neurons with partially correlated inputs can be measured to obtain correlation gain[33,35,40]. One should note that for many of these studies, such as correlation gain or dynamic gain measurements, only the conductance waveform and not the absolute scale of conductance fluctuations needs to be known.

CoDyPs may also turn out effective for controlling the activity of intact networks *in-vivo*. Modeling studies of cortical networks raise the possibility that driving only a subset of neurons with naturalistic inputs can effectively control the state of the entire network if the inputs are shaped to match network-generated inputs[41]. While more theoretical work is needed to clarify the dynamic properties of cortical networks[36,42,43], one expects in general that complex and time dependent inputs can control the network dynamics while preserving its intrinsic complexity[44]. CoDyPs can be used to examine whether such naturalistic perturbation approaches can be used to control cortical networks *in-vivo*.

Combining CoDyPs with approaches to simultaneously detect action potentials over long periods of time from many individual neurons in parallel will enable to address new questions in cellular physiology of neuronal computation. Screening for the effect of mutations or short term protein knockdown will allow the dissection of the protein network underlying dynamical properties of neurons. Comparisons of individual neurons might reveal individual differences underlying distinct dynamical properties. Combined with patch-clamp measurements in previously characterized cells it will be possible to reveal the biophysical basis of encoding diversity.

## Methods

*Cell culture and transfection*

A monoclonal stable cell line of HEK 293 cells expressing Channelrhodopsin-2 tagged with the fluorescent marker YFP was established. The pcDNA 3.1- Chop2-YFP construct was kindly provided by Ernst Bamberg, (MPI for Biophysics, Frankfurt, Germany). The construct ChIEF-tdTomato was kindly provided by Dr. Roger Y. Tsien (UCSD). The transfection of the HEK 293 cells with ChIEF-tdTomato was performed with Amaxa Nucleofector (Amaxa), using $10^6$ cells, 4 µg of DNA and program A-23. Cells were cultured in DMEM/F12 + GlutaMaxTM medium (Gibco 31331-028) supplemented with 10% Fetal Calf Serum. For the stable cell line Geneticin (100 mg/ml) was added to maintain the selection pressure. The cells were plated on poly-L-lysine coated coverslips for electrophysiological recordings.

*Western blotting*

Standard protocols were followed[45]. In brief, the protein was extracted from the cells (wild type HEK and HEK Chop2 cell line) using 200 µl lysis buffer. The cell suspension was mixed with 7.5 µl loading buffer and 3 µl reducing agent, heated for 10 minutes at 70 °C. The sample was then loaded in NuPAGE Novex Bis-Tris Mini Gels (Invitrogen) and run for 35 minutes in the running buffer at 200 V. Then the blotting procedure was performed. The running settings of the blot were as follows: 10 V, 20 V and 30 V for 10 minutes each, followed by 40 V for 20 minutes and 50 V for one hour. After running, the membrane was blocked for 1 hour in blocking solution and then incubated in 10 ml blocking solution with the primary antibody anti – GFP (abcam, Germany; $10^{-3}$ dilution) overnight at 4 °C with shaking then incubated in 10 ml blocking solution with 1 µl of secondary anti-rabbit antibody ($10^{-5}$ dilution) for one hour at room temperature. Then the blot was developed. The band of interest is around 60 kDa.

*Electrophysiology*

The HEK Chop2 cells were identified with the YFP fluorescence and the ChIEF transfected HEK cells were identified with td tomato fluorescence under an inverted microscope (Axiovert 135, Zeiss). Patch pipettes were fabricated from PG10165-4 glass (World Precision Instruments) using a vertical puller (L/M-3P-A, List Medical). Pipette resistance typically lay between 2 and 4 MOhm, yielding access resistances between 3 and 8 MOhm. In the few cases when this range was exceeded, series resistance compensation was used. The extracellular solution consisted of (in mM) 145 NaCl, 3 KCl, 2 $CaCl_2$, 1 $MgCl_2$, 10 HEPES, and 20 glucose (pH 7.35, 290 mOsm). It was supplied to the recording chamber at a rate of about 0.5 ml per minute. The intracellular solution consisted of (in mM) 110 NaCl, 10 $Na_2EGTA$, 4 $MgCl_2$ and 10 HEPES (pH 7.4, 300 mOsm). Immediately before the patch pipette touched the cell the microscope objective was replaced by the light source mounted on the objective revolver (described below). Whole cell patch clamp was performed using an EPC9 amplifier controlled by Patchmaster (both HEKA Elektronik). The membrane potential was -60 mV throughout. Currents were low-pass filtered at 2 kHz (4-pole Bessel filter) and sampled at 10 kHz. Using Patcher's Power Tools (Dr. Francisco Mendez, Frank Würriehausen) the current recordings were imported into Igor Pro (Wavemetrics) and analyzed automatically with custom written protocols.

*Multielectrode array recordings*

Hippocampal neurons isolated from E18 rats were cultured on multielectrode arrays (MEA; type 200-30iR from Multichannel Systems) coated with a mixture of poly-D-lysine and laminin at a density of 1000 cells per mm². The cells were kept in an incubator at 37°C and a mixture of 5% $CO_2$ + 95% $O_2$. The cultures were transfected after 14 days *in-vitro* with AAV-SynapsinI-CHOP2; recordings were done after 21 days *in-vitro*. During the long-term recordings, cultures were perfused with Neurobasal/B27 medium (Invitrogen) with fibroblast growth factor (FGF). Synaptic blockers against AMPA, GABA-a and NMDA receptors were applied from 30 minutes before start of the long-term illumination (10 µM NBQX, 1.5 µM Picrotoxin and 50 µM APV); this completely abolished action potentials in the absence of light stimulation. The perfusion solution was continuously bubbled with a mixture of 5% $CO_2$ + 95% $O_2$. Data from MEAs were captured at 25kHz using a 64-channel A/D converter and MC_Rack software (Multichannel Systems). Amplifier gain was set to 1100. After high pass filtering (100 Hz) action potentials are detected in a cutout recorded 1ms before and 2ms after crossing a threshold of -40µV, which was > 3 standard deviations of the baseline activity.

*Light source*

The key requirements for the light source were high light power at around 480 nm, fast and well controllable modulation of the light power and stable illumination over several hours. An additional requirement for the illumination of the spatially extended MEAs is homogeneous light-power density over an area of $1 \times 1$ mm. All these requirements were met by a blue light emitting diode (LED, Luxeon rebel color with Lambertian dome, Philips Lumileds) with 5 W maximal power consumption, placed 25 mm below the illuminated cells. The light output was controlled via the voltage of the D/A-board of the EPC-9 patch clamp amplifier, converted to current in a custom made analog driver circuit, resulting in a input of 1 W at the LED for each Volt at the D/A-board. Rise-time to maximum Light power was < 20 µs. Step protocols of light-power density were created in Patchmaster's pulse protocol editor, fluctuating light stimuli were synthesized in Matlab and played through Patchmaster.

The LED was mounted inside a metal cylinder, closed except for a 5 mm aperture, 8 mm above the LED's active surface. The metal cylinder was screwed in a slot of the objective revolver of the inverted microscope, securing a highly reproducible and fast placement of the LED on the optical axis of the microscope. At the same time the cylinder completely prevented stray signals of the power line supplying the LED from contaminating the patch clamp signal. In this configuration the maximum output of the LED equated to 0.27 mW/mm$^2$ in the plane of the cells. The voltage-light relationship of the driver and the LED was slightly sublinear, possibly due to a non-ideal behavior of the driver, combined with a small temperature effect of the LED. In the experiments on MEAs, a photodiode, included in the LED housing recorded the applied light intensity at all times.

*Photometry*

The irradiance at the level of the cells is measured with an ordinary Si photodiode, placed in an adjustable, baffled holder which allows exact positioning relative to the light source and admits no stray light to the active area of the diode. The geometry of the radiation pattern of the LED is not altered by any optical components, so that a photodiode with a small active area (Hamamatsu S2386-18K) can just be placed in the location of the object to be illuminated, and the incident light power arriving at this diode can be measured. This directly yields a value for the irradiation averaged over the area of the diode. In order to reduce the incident light power, a neutral density filter is first calibrated in a control measurement and then placed in front of the entrance aperture of the photodiode holder.

The photodiode is connected to a current-to-voltage amplifier (NEVA7212) the output signal of which is displayed on an oscilloscope against the periodic driving voltage used to control the power amplifier circuit, which, in turn, drives the LED. This directly yields the characteristic of the LED and its driving circuit. Further details can be found in the supplementary methods.

*Fluctuating stimuli*

Synthesis of a time series $\{V_i\}$ of command voltages with a time step $\Delta t$ followed the iterative rule:

$$V_i = -(V_{i-1} - \overline{V}) \cdot \kappa + \sigma\sqrt{1-\kappa^2}\xi_i, \quad \text{with} \quad \kappa = \exp(-\Delta t / \tau_{corr}) \qquad (1)$$

where the $\xi_i$ are provided by a generator of $N(0,1)$ normally distributed random numbers. Equation 1 generates an Ornstein-Uhlenbeck process with the time average $\overline{V}$, the variance $\sigma$ and the correlation time $\tau_{corr}$. To protect the LED the voltage sequence was restricted to lie between 0 and 5 V. An alternative way to construct the sequence $\{V_i\}$ is to pass the white noise $\{\overline{V}+\sigma\cdot(1+\kappa)^{0.5}\cdot(1-\kappa)^{-0.5}\cdot\xi_i\}$ through an RC-type low pass filter with the time constant $\tau_{corr}$. The light stimuli used here were all synthesized with the same random number sequence $\{\xi_i\}$ and represent therefore just differently filtered versions of the same white noise sequence. Three different correlation times were used: 1 ms, 5 ms and 50 ms. For each $\tau_{corr}$ four different combinations of average and standard deviation (SD) were generated, referred to as condition c1 to c4. Considering the slightly non-linear current-light relationship of the photo-diode, the respective values of average and standard deviation of the light-power were (in mW/mm²): (c1) 0.134 and 0.057; (c2) 0.161 and 0.052; (c3) 0.185 and 0.046 and (c4) 0.177 and 0.068. Each of the twelve different stimuli was presented ten times. In a subset of experiments (n=8) this was done in random order. In other cells the stimuli were presented in an interleaved order.

*Data Analysis*

Current responses to light steps were described by single exponential functions. As the currents deviate from a single exponential time course by a slight delay at the onset (< 1 ms) and, especially for ChR2, by the inactivation, the choice of the range to be fitted does influence the results of the exponential fit. This influence in minimized by starting the fit only 1 ms after the onset of the light step and by choosing the fit range's duration according the estimated time constant. To this end the fit was iterated and after each round the fit range was set to three times the estimated time constant, until the change in this estimate was smaller than 3 %.

Each of the twelve different fluctuating light stimuli (four conditions c1 to c4 and three correlation times) was presented ten times to a given cell. The sequence of presentation was interleaved and in a subset of experiments (n=8) it was random. The ten current responses of a given cell, which were elicited by the individual trials with a given stimulus, were averaged to give the average response of this cell to this stimulus. When stimuli were applied in random order, the time between the first and last trial of a given stimulus was as long as 10 minutes. During this time small changes in the recording conditions or in the leak current could occur, causing a small offset between the respective currents. To demonstrate the reproducibility of the light-induced component of the measured currents, the offsets were accounted for by shifting the individual responses to achieve the same trial average for all trials. This was done only for the display in Fig. 3 but not for quantification. Power spectral density was calculated over 50 % overlapping intervals of 409.6 ms duration (4096 points), windowed with a Welch function. Before averaging, power spectral density of different recordings was normalized to 1 at 7.3 Hz, the frequency bin with the maximum power. This normalization assured that the shape of the average power spectral density faithfully reflects the average shape of all individual examples.

Pearson correlation coefficients $r_P$ were calculated for successive trials of a given stimulus in a given cell. Those nine individual $r_P$ values were averaged to give the average $r_P$ for this cell and stimulus. The average autocorrelation function for a given cell and stimulus was calculated from the average response of this cell and stimulus. Autocorrelation functions were normalized to 1 at t=0. The average impulse response function (IRF) for a given cell and stimulus was calculated as the inverse Fourier-transformation $F^{-1}$ of the transfer function, which is the ratio of the Fourier-transform of the average response I(t) of this cell to this current $F(\,I(t)\,)$ and the Fourier - transform of the respective stimulus $F(\,S(t)\,)$:

$$IRF(t) = F^{-1}\left(\frac{F(I(t))}{F(S(t))}\right). \qquad (2)$$

The overall averages of the autocorrelation function and IRF for a given stimulus were calculated as average over the respective cell averages. As the amplitudes of the IRFs varied considerably between cells and because the important aspect is the shape of the IRF more than its amplitude, IRF were normalized by their integral before averaging. The resulting average was then multiplied by the average integral of the individual IRFs to reveal a representative average IRF shape and amplitude. The 95% confidence intervals of the average autocorrelation functions and IRFs were computed by balanced bootstrap: the averages from N cells were each cloned 1000 times to yield N×1000 traces. Those were randomly grouped in 1000 samples of N traces each. Each sample was averaged resulting in 1000 bootstrap averages. For each time point the lowest 25 and largest 25 values of all the bootstrap averages are identified. The range covered by the remaining 950 values represents the bootstrap confidence interval at this time point.

The auto-correlation functions were fit with

$$C(\tau) = \frac{\tau_{corr} + \tau_{cut-off}}{2} \left( \frac{\exp\left(-\frac{|\tau|}{\tau_{corr}}\right) + \exp\left(-\frac{|\tau|}{\tau_{cut-off}}\right)}{\tau_{corr} + \tau_{cut-off}} + \frac{\exp\left(-\frac{|\tau|}{\tau_{corr}}\right) - \exp\left(-\frac{|\tau|}{\tau_{cut-off}}\right)}{\tau_{corr} - \tau_{cut-off}} \right). \quad (3)$$

This equation describes normalized the auto-correlation of an Ornstein-Uhlenbeck process with correlation time $\tau_{corr}$ passed through a RC-type low-pass filter with the time constant $\tau_{cut\text{-}off}$ (Reference Mins Arbeit?). The IRFs were described with a function comprising an initial delay $t_d$, followed by an exponentially growing term multiplied with an exponentially decaying term:

$$IRF(t) = \begin{cases} 0 & \forall t < t_d \\ A \cdot \left(1 - \exp\left(-\frac{t - t_d}{\tau_{act}}\right)\right) \cdot \exp\left(-\frac{t - t_d}{\tau_{inact}}\right) & \forall t \geq t_d \end{cases}. \quad (4)$$

As the light activated at current -60 mV has a negative sign, the amplitude A is negative.

*Estimating relative ion fluxes*

We use the Goldmann-Hodgkin-Katz flux equation:

$$\Phi_S = P_S z_S^2 \frac{V_m F^2}{RT} \frac{[S]_i - [S]_o \exp(-z_S V_m F / RT)}{1 - \exp(-z_S V_m F / RT)}. \quad (5)$$

Here F, R, T and $V_m$ have the standard meanings of Faraday's constant, gas constant, temperature and membrane voltage. The abbreviations $\Phi_S$, $P_S$ and $z_S$ denote the flux, permeability and valence of ion species S. With a membrane voltage of -60 mV the exponential term equals 10.8. For monovalent ion species S1 and S2 this simplifies to a flux ratio:

$$\frac{\Phi_{S1}}{\Phi_{S2}} = \frac{P_{S1}}{P_{S2}} \frac{[S1]_i - [S1]_o \cdot 10.8}{[S2]_i - [S2]_o \cdot 10.8}. \quad (6)$$

Using typical ion concentrations (in mol/l $[Na]_i=0.005$, $[Na]_o=0.15$, $[K]_i=0.15$, $[K]_o=0.003$, $[H]_i=[H]_o= 10^{-7}$),

## Acknowledgments

We thank Roger Tsien and John Lin (Stanford) for the ChIEF Plasmid. Tatjana Tchumatchenko kindly provided voltage traces for fig.1. This work is supported by the German Ministry for Education and Research (BFNT grants 01GQ0811 and 01GQ0813; BCCN II grants 01GQ1005B and 01GQ1005E).

*Author contributions*

FW, WS, TG and EB conceived the project, FW, WS, AN, AEH, KB and RF designed experiments; AN, AEH and JN performed experiments with assistance of KB; OMS generated the AAV ChR2 virus; AN, AEH, FW, JN analyzed the data with assistance of GA; AN, AEH and FW supervised experiments and wrote the manuscript.

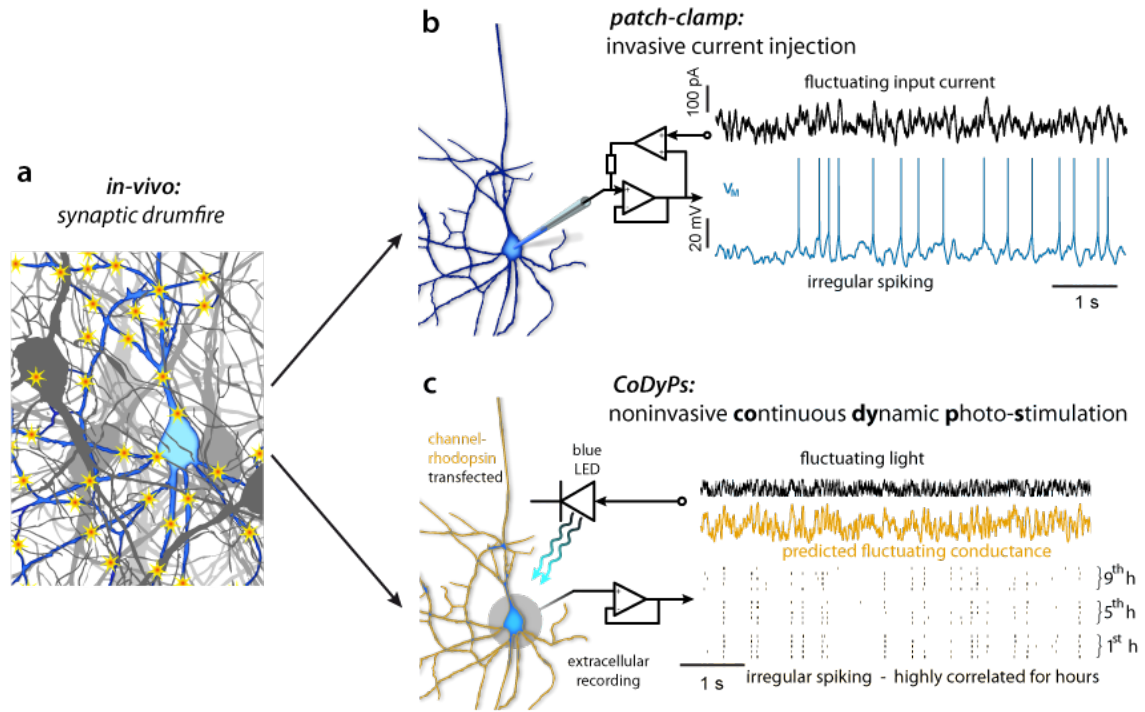

**Figure 1: Two ways to study *in*-vivo-like fluctuation driven spiking activity under controlled conditions**

**a,** schematic representation of the ongoing synaptic drumfire to which neurons in the CNS are typically exposed. Sparks represent active synapses. Cortical pyramidal neurons will typically receive synaptic inputs at a rate of several kilohertz. **b and c,** two alternative experimental approaches to emulate the resulting input fluctuations and register the fluctuation driven activity *in-vitro*: whole cell current injection (b) and CoDyPs (c), here depicted for a neuron cultured on a circular extracellular electrode. In contrast to the whole cell stimulation/recording, CoDyPs offers extended recording and stimulation/recording of multiple neurons simultaneously.

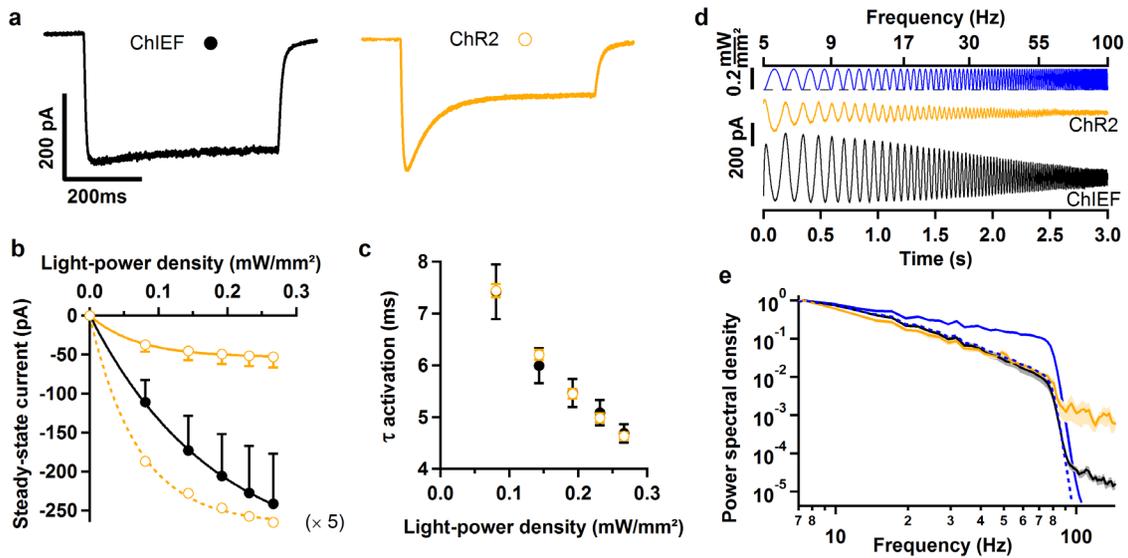

**Figure 2: ChIEF supports large steady-state currents with a low-pass filter characteristics similar to ChR2**

**a**, Representative current responses to 1 s light stimuli at 0.27 mW/mm² recorded at a membrane potential of -60 mV from HEK 293 cells. Orange: stable cell-line expressing ChR2, black: transiently expressing ChIEF. **b,** Relationship between steady-state current and light-power density for ChR2 (orange, n=8) and ChIEF (black, n=11) (error bars indicate standard error). The straight lines are single exponential fits. A five times scaled up version of the ChR2 data (dashed) is given to highlight the difference between the two Channelrhodopsin variants. **c,** The activation time constants (see Material and Methods) of ChIEF and ChR2 currents are equally dependent on the light-power density. **d,** A chirp stimulus (blue; frequency 5 to 100 Hz, see upper axis) evokes current responses with decreasing modulation depth, indicating the low pass behavior of the light-activated currents. Representative current responses to the chirp stimulus are shown in orange (ChR2) and black (ChIEF). **e,** Average normalized power spectral density of responses from ChR2 (orange) and ChIEF (black) are nearly identical. The power spectral densities of the light stimulus (continuous blue line) and a low-pass filtered version of the light stimulus (dashed blue line, -3 dB cut-off frequency of 20 Hz) are displayed for comparison. Standard errors are shown as brighter bands. Both channelrhodopsin variants transform the power spectrum similar to the single pole low pass filter.

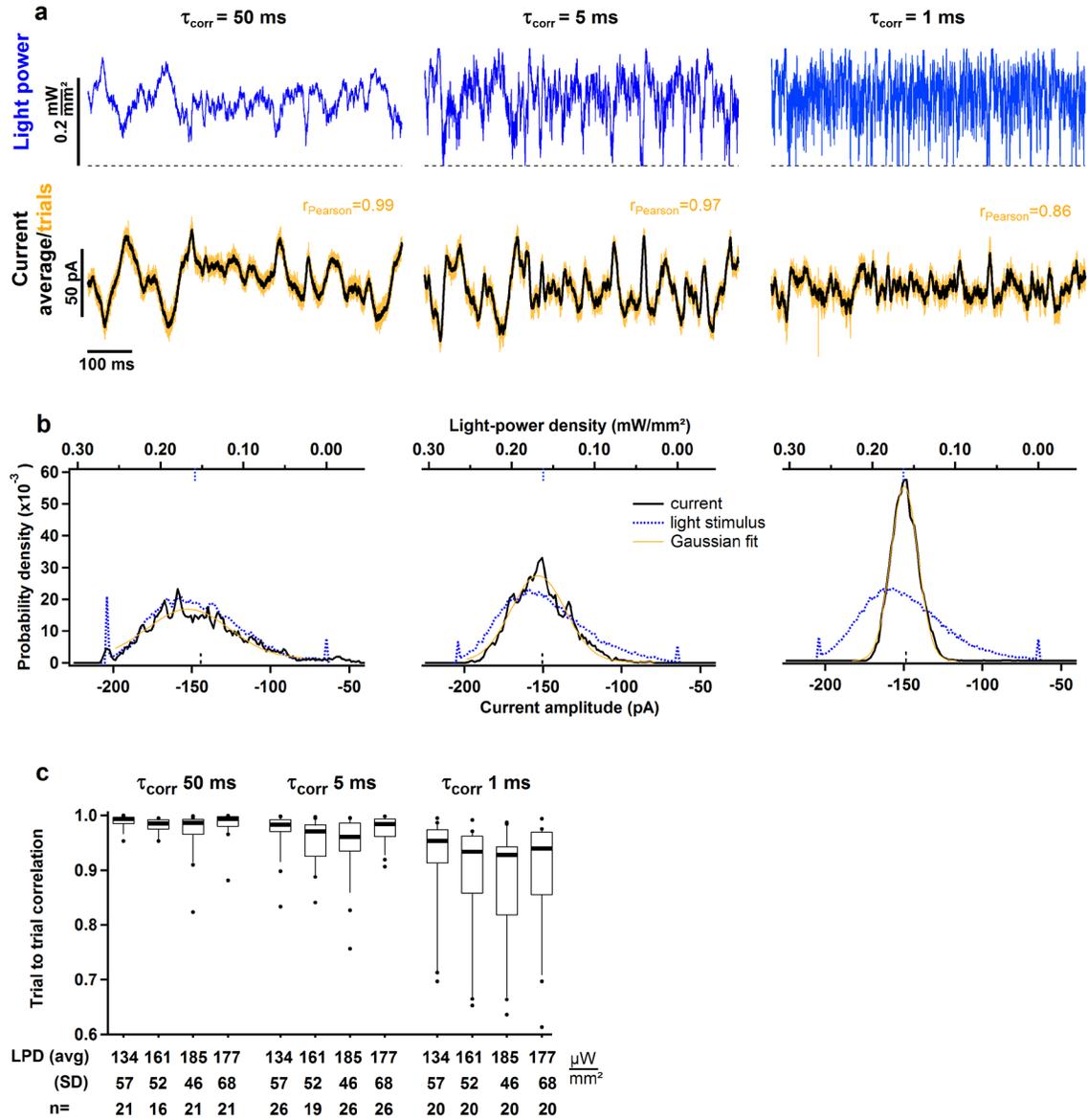

**Figure 3: Trial to trial reproducibility of CoDyPs driven currents**

**a,** 700 ms periods from a representative HEK 293 cell expressing ChIEF. All light stimuli (blue) stem from the same realization of a random Ornstein-Uhlenbeck process (condition c2, see Methods). Stimuli only differ in correlation time $\tau_{corr}$, decreasing from 50 ms (left) to 1 ms (right). Ten individual current traces, driven by the light stimuli, are displayed (orange), average currents are superimposed (black). The evoked current is highly reproducible, indicated by the high average correlation coefficients of successive current traces (see also **c**). The dotted line represents zero for both: light stimulus and light activated current response. Note how for $\tau_{corr}$=50 ms the response essentially mirrors the light stimulus. For shorter correlation times the amplitude of the driven current excursions decreases as the response no longer follows the faster signal amplitude modulation. The residual differences between individual responses and average have a standard deviation of 4 to 5 pA. **b,** Histograms of current amplitude (black) and stimulus amplitude (blue, dotted) are displayed in a tentative alignment. Average values are indicated as short vertical lines at the respective axis. While the stimuli have, by design, nearly identical histograms for all correlation times, the current amplitude histograms narrow as the correlation time decreases.

**c,** Box-plot of the trial to trial Pearson correlation coefficients. Data are grouped by illumination condition, the average light-power density and the standard deviation are given in µW/mm² Median is indicated by the black bar, the box comprises the central 50 percent of points and the whiskers the central 80 percent. Individual points represent outliers in the lowest and highest 10 percent. The prominent outliers for short correlation time (1 ms) stem from four cells with little light induced current (<100 pA) and a leak current of the same order of magnitude. As the light driven current excursions are larger for longer correlation times the correlation coefficients from the same cells are larger for $\tau_{corr}$ of 5 and 10 ms.

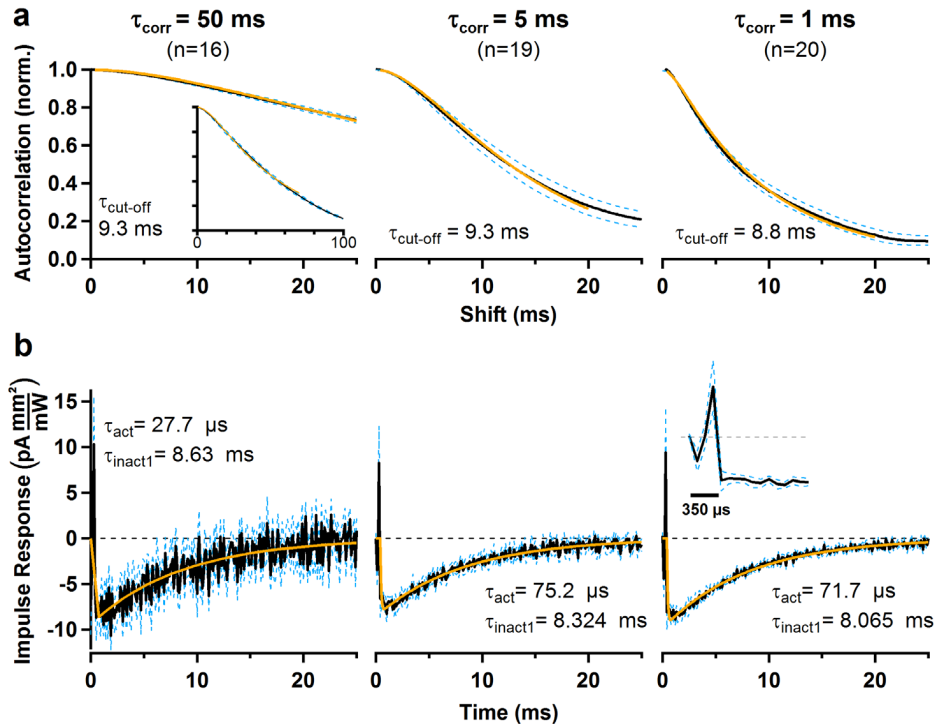

**Figure 4: The statistics of CoDyPs driven fluctuating currents obeys linear response theory,**

**a,** The normalized autocorrelation functions (black) conform with the prediction (orange) for an Ornstein-Uhlenbeck process, low-pass with cut-off time constant $\tau_{\text{cut-off}}$ (equation 3). Shown here are results for condition c2 (light-power density 161 µW/mm² average, 52 µW/mm² standard deviation).
**b,** Average impulse response functions (black lines) derived from ChIEF mediated currents activated by fluctuating light stimuli in condition c2. The temporal structure, the $\tau_{\text{corr}}$ of the stimuli has no influence on the shape of the impulse response function. Following an initial transient (inset for $\tau_{\text{corr}}$=1 ms, see Results) and a rapid onset, the impulse responses are well described by a single exponential function (orange)
The dashed light blue lines in a and b enclose 95 % bootstrap confidence intervals.

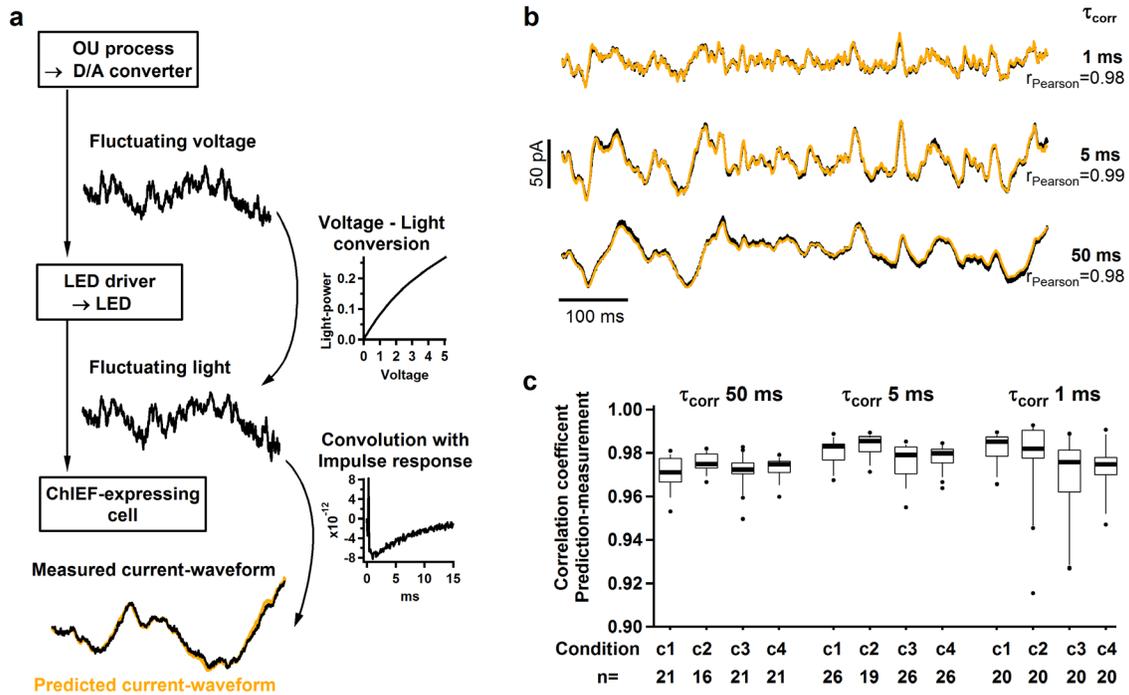

**Figure 5: Computational prediction of CoDyPs-driven currents**

**a,** flow chart depicting the prediction of CoDyPs induced currents: a fluctuating voltage signal is fed through the digital/analog board to the LED-driver. By means of the transfer function of the LED-driver the light waveform can be calculated. This is folded with the IRF of the employed channelrhodopsin to obtain the predicted conductance chance.
**b,** The average current responses from Fig. 3 are shown in black, vertically displaced for clarity. Response predictions, constructed by convolution of light stimuli (Fig. 3 top panels) and average impulse response functions (a) are overlaid in orange. These predictions closely match the actual currents after they have been scaled and offset according to the mean and standard deviation of the current waveform. For $\tau_{corr}$= 50 ms the noisy average impulse response function was substituted for by the fit function (grey in a) to reduce the noise level in the prediction. **c,** For each correlation time the coefficient of correlation between the prediction and the average current is very high.

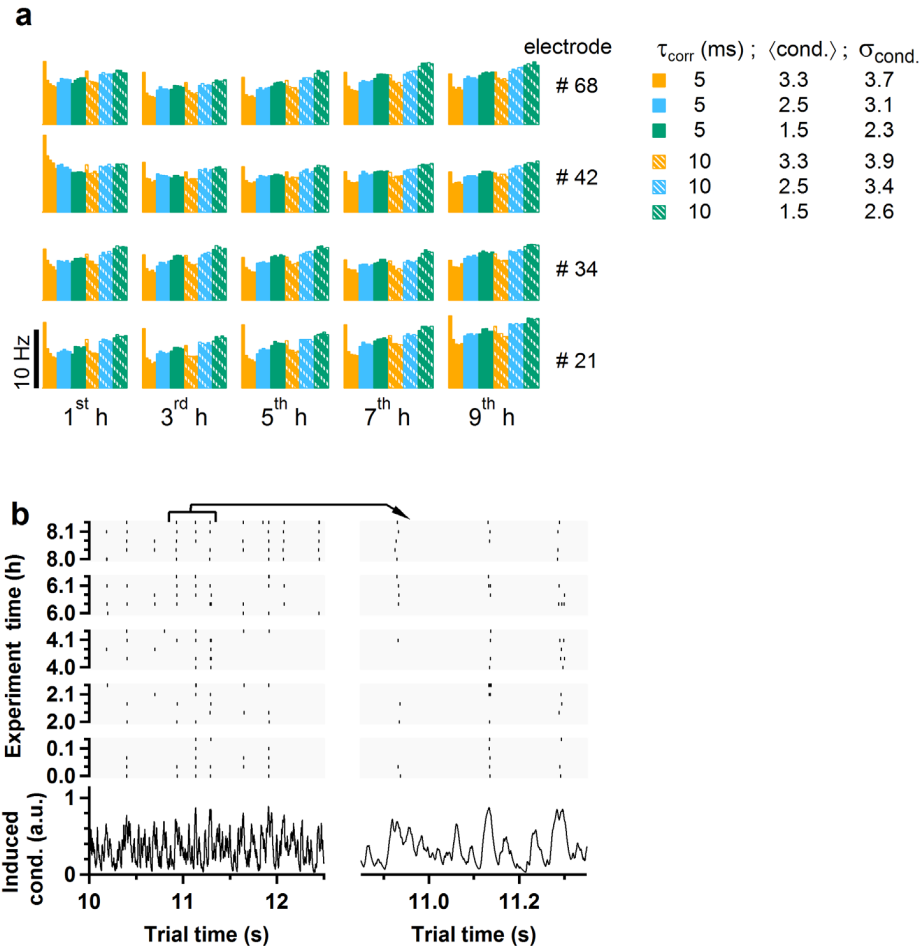

**Figure 6: CoDyPs elicits stable and highly correlated action potential sequences over many hours**

**a,** Neurons expressing ChR2 were cultured on multi-electrode arrays, permitting non-invasive detection of action potentials. detected by extracellular Each of six different 2 minute light stimuli was presented five times in a row, totalling 60 minutes of stimulation. For the 4 electrodes with the highest firing rates the average rate of action potentials is plotted for each 2 minute stimulation period. The conductance predicted for the stimuli differed by average and standard deviation (3 different levels each) and the correlation time (5 and 10 ms). The 60 minute block was repeated five times interspersed by one hour darkness. Changing the stimulus reproducibly changes the action potential rate. Onset of stimulation after 1 h darkness causes a very strong transient increase in the firing rate **b,** Raster plots of spike times, displayed above the predicted light induced conductance waveform, show that spike patterns were stable and highly correlated over many hours

*Continuous Dynamic Photostimulation - injecting defined, in-vivo-like fluctuating conductances with Channelrhodopsins*

Andreas Neef *, Ahmed El Hady*, Jatin Nagpal*, Kai Bröking, Ghazaleh Afshar, Oliver M Schlüter, Theo Geisel, Ernst Bamberg, Ragnar Fleischmann, Walter Stühmer, Fred Wolf

## Supplementary Results:

*Channelrhodopsin 2 has similar dynamic properties as ChIEF*

In a small number of HEK-cells, stably transfected with ChR2, the currents induced by fluctuating light (average intensity: 161 µW/mm², standard deviation: 52 µW/mm² $\tau_{corr}$=5 ms) were analyzed to obtain an estimate for the auto-correlation and the impulse response function under these conditions. The results were similar to those obtained from ChIEF transfected cells (Fig. S1)

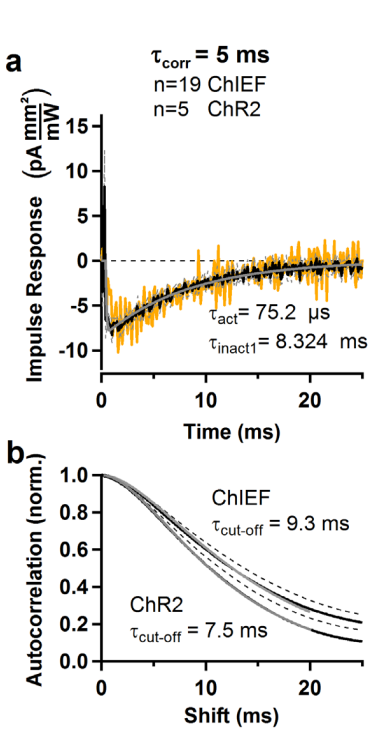

**Supplemental figure 1: ChR2 and ChIEF have similar response characteristics.**

**a,** The impulse response function of ChR2 has a similar shape but a smaller amplitude. The red trace represents x7 scaled impulse response function of ChR2. The amplitude is much smaller than for ChIEF because the steady state current amplitude is much smaller. The shape of the impulse response function is very similar.

**b,** The autocorrelation function of ChR2 can be well described by equation 4. The cut-off time constant, the only free parameter, was estimated to be slightly smaller than for ChIEF.

*Additional examples of action potential sequences under CoDyPs*

**Supplemental figure 2: More examples of spike patterns from long-term experiments**
**a,** the probability density functions of the 6 different stimulus ensembles. **b,** fluorescence image of the multi-electrode array, the tdTomato label of the ChR2 is shown cyan. Dark circles are extracellular electrodes. **c-f,** raster plots from the electrodes 23, 33, 43, & 73 covering 3 s. Scaling see bottom left.

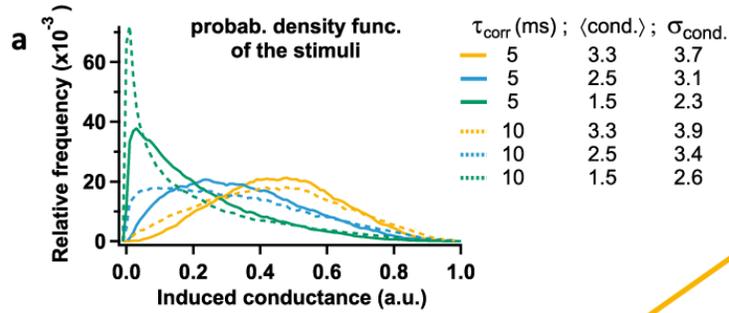
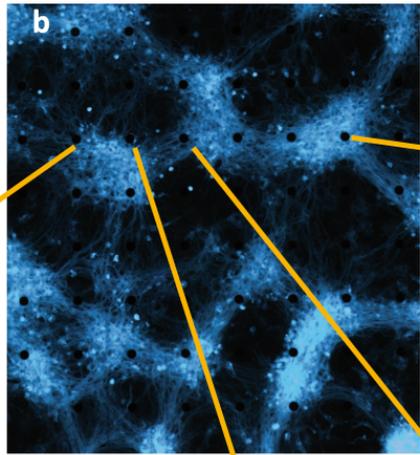
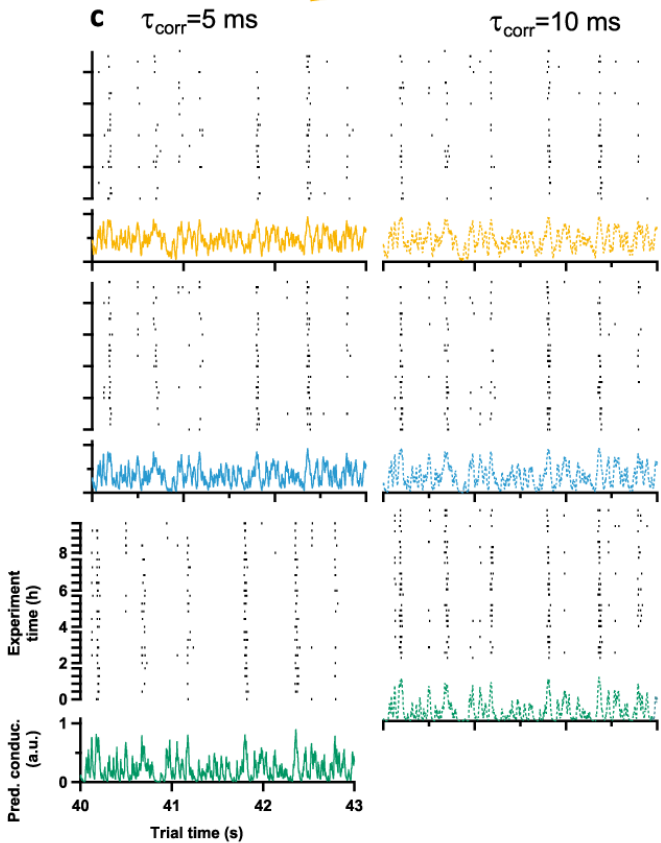
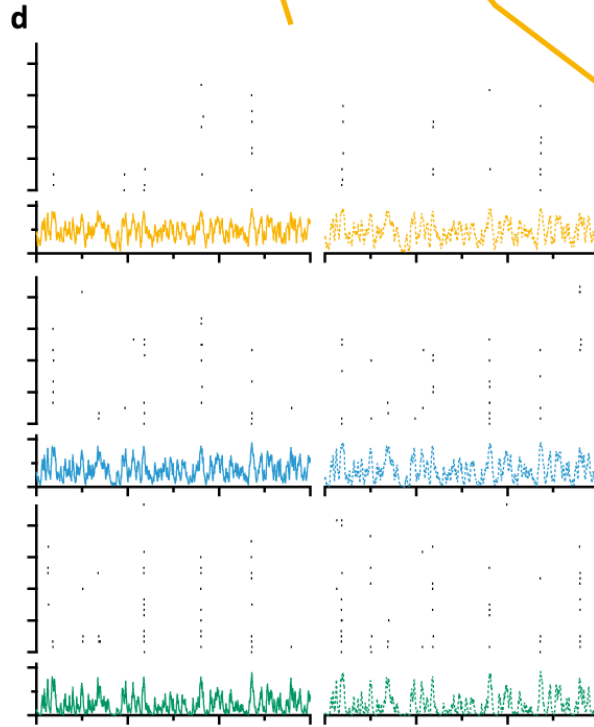
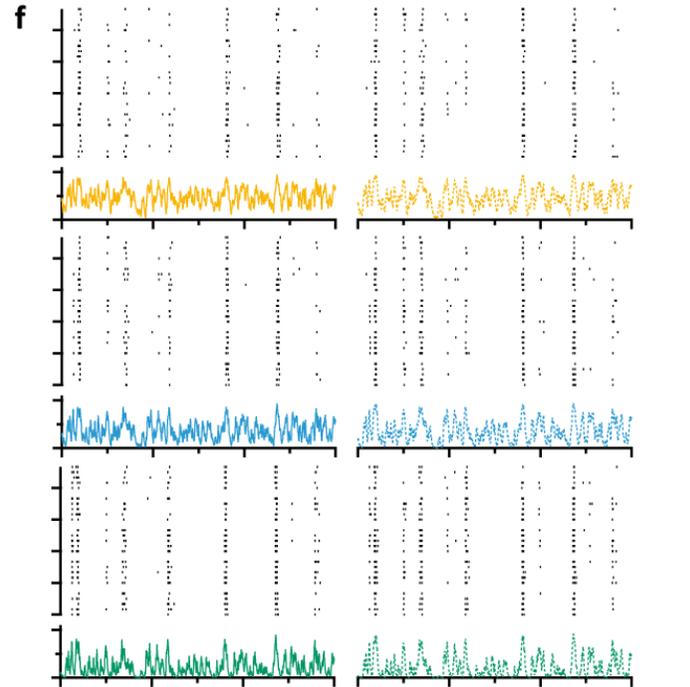
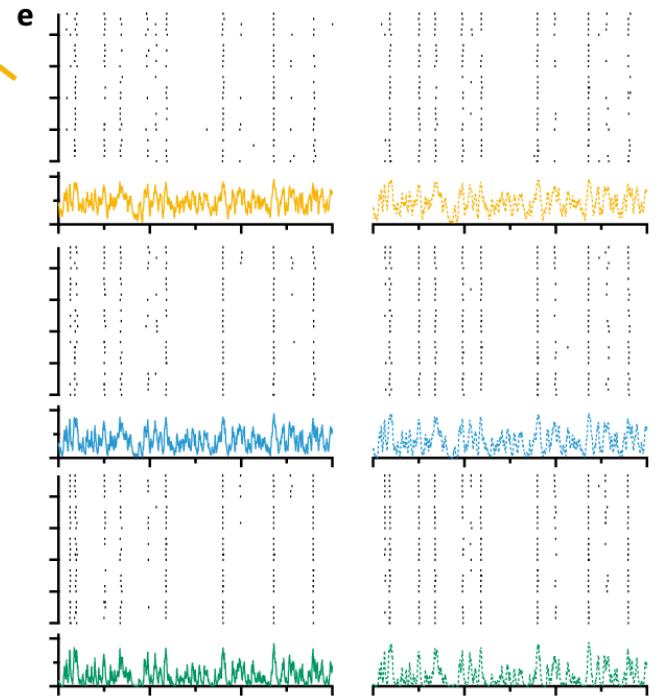

Supplementary Methods:

*Methods: Photometry*

An ordinary Si photodiode measures the irradiance, the cells are subjected to. Since no optical components are in use to alter the geometry of the radiation pattern of the LED, it is sufficient to just place a photodiode with a small active area in the location of the object to be illuminated, and to measure the incident light power arriving at this diode. This procedure yields a value for the irradiation averaged over the area of the diode (in our case 1.21 mm$^2$).

The spectral properties of the light source had been determined beforehand with a small fiber-coupled spectroscope: The spectrum of the Luxeon LED turns out to be almost exactly symmetric and narrow, so that for all calculations the spectral response for the spectral peak at 480 nm, namely 0.276 A/W can be used. The photodiode employed, a Hamamatsu S2386-18K, was checked against a calibrated diode of a similar type (Hamamatsu S2386-8K) in a control measurement. For convenient handling, it was mounted in the photodiode holder from a Zeiss ElDi2 military rangefinder. This holder consisted of a brass tube of 6 mm internal diameter and 8 mm of length, to the front of which a thin disk of stainless steel with a hole of 3.1 mm diameter was glued. The latter serves as a baffle to prevent the wall of the tube from being illuminated, which would send stray light to the diode. Additionally, the inside of the tube was thoroughly blackened with soot. The efficiency of the baffling was checked by gradually pointing the holder towards a bright light source in an otherwise dark environment: Only when the light shone straight onto the detector the photocurrent was detected, otherwise, no measurable current was observed.

The diode itself sits in a small ceramic tube and can be shifted in a plane perpendicular to the axis of the holder with four adjustment screws to ensure correct position. After the position of the diode relative to the aperture was measured, we aligned the diode and measured its position along the axis of an optical bench, on which it is mounted at a known distance from the high power LED. The photodiode was connected to a current-to-voltage amplifier (NEVA7212) the output signal of which is displayed on an oscilloscope against the driving voltage used to control the power amplifier circuit, which, in turn, drives the LED. For the measurement, a delta voltage signal at a frequency of 20 Hz is used. This directly yields the characteristic of the LED and its driving circuit.

Since the high irradiance would otherwise saturate the photodiode, the irradiance deriving from the LED must be dampened by a known factor in order to make any measurement at all. This can be achieved by two methods: First, a neutral density filter was used, that was calibrated for the spectral range in question. As a second, more precise method we placed the detector at a distance *r* from the light source and exploit the fact that the irradiance drops as *1/r$^2$* as the distance increases. Placing the filter between the LED and the photodiode, and putting the light source at a position $r_1$ from the detector, the current through the photodiode was measured. The filter was then removed and the detector moved to a distance $r_2$ from the light source at which an identical photocurrent was observed. The neutral density of the filter in question then is

$$ND = \lg\left(\frac{r_2^{\,2}}{r_1^{\,2}}\right).$$

Having thus measured the neutral density of the filter for the spectral range in question, it can be used for further measurements in situ. Since the whole spectrum of the LED lies between 460 nm and 500 nm, a disk of thermal protection glass by Schott was used as a filter, which is then calibrated for neutral density.

The filter was placed in front of the aperture of the detector. With the irradiance diminished in this way, the incident light power at 25 mm distance from the emitter of the LED can be calculated from a measurement of the photocurrent through the diode. Control measurements using the first method, viz. placing the detector at a distance from the source, yielded the same results to within a small error, which may be due to reflective losses at the front and back side of the ND filter.